\documentclass{ws-procs9x6}

\begin{document}

\title{Infrared QCD and the Renormalisation Group\footnote{\uppercase{P}resented by \uppercase{DFL} at \uppercase{S}trong and \uppercase{E}lectroweak \uppercase{M}atter, \uppercase{H}elsinki, \uppercase{F}inland, \uppercase{J}une 2004. \uppercase{W}ork supported by \uppercase{EPSRC} and by the \uppercase{DFG} contract \uppercase{SM}70/1-1.}}

\author{Daniel F.~Litim}

\address{CERN, Theory Group, CH -- 1211 Geneva 23, and\\
School of Physics, University of Southampton, Southampton SO17 1BJ, U.K.}

\author{Jan M.~Pawlowski\footnote{\uppercase{P}resent address: \uppercase{I}nst.~f.~\uppercase{T}heor.~\uppercase{P}hysik, \uppercase{P}hilosophenweg 16, 69120 \uppercase{H}eidelberg, \uppercase{G}ermany.}\, , Sergei Nedelko\footnote{\uppercase{P}resent address: \uppercase{BLTP}, \uppercase{JINR}, 141980 \uppercase{D}ubna, \uppercase{R}ussia.}\,, Lorenz v. Smekal}

\address{Institut~f\"ur~Theoretische~Physik III, 
Universit\"at Erlangen,\\ 
Staudtstr.~7, 91058 Erlangen, Germany}

\maketitle

\abstracts{ We study the infrared regime of QCD by means of a Wilsonian
    renormalisation group. We explain how, in general, the infrared structure
    of Green functions is deduced in this approach. Our reasoning is put to
    work in Landau gauge QCD, where the leading infrared terms of the
    propagators are computed. The results support the Kugo-Ojima scenario of
    confinement. Possible extensions are indicated.}


\section{Introduction}
Many aspects of QCD are well understood in its ultraviolet limit where the
gauge coupling is small as a consequence of asymptotic freedom. Therefore,
perturbation theory is expected to provide a viable description of QCD
phenomena at high energies or high temperature.  On the other hand, in the
infrared limit, quarks and gluons are confined to hadronic states and the
gauge coupling is expected to grow large. In consequence, a reliable
description of low energy phenomena -like confinement, the physics of bound
states, chiral symmetry breaking or the confinement-deconfinement transition-
call for a non-perturbative analysis.

Renormalisation group methods provide important analytical tools in
the study of non-perturbative phenomena. In this contribution, we
study Landau gauge QCD in the non-perturbative infrared
regime~\cite{Pawlowski:2003hq} by means of an exact (or functional)
renormalisation group \cite{Wegner,Wetterich:yh}.  The strength of the
approach is its flexibility when it comes to
approximations. Furthermore, efficient optimisation procedures are
available, increasing the domain of validity and the convergence of
the flow.\cite{Litim:2001up} We discuss how, in general, the
momentum structure of Green functions in the infrared is deduced from
non-perturbative renormalisation group equations.  As an application
we compute the leading non-perturbative infrared coefficients (or
anomalous dimensions) for the gluon and ghost propagator in Landau
gauge. Our results are discussed in the light of the Kugo-Ojima
confinement criterion and earlier findings based on other
non-perturbative methods. For technical details we refer to the
original publication.\cite{Pawlowski:2003hq}

\section{Renormalisation group for QCD}

The exact renormalisation group is based on the Wilsonian idea of
integrating-out momentum degrees of freedom within a path integral
representation of quantum field theory. Central to this approach is the
effective action $\Gamma_k$, where quantum fluctuations with momenta $q^2>k^2$
are already integrated out. The renormalisation group equation for $\Gamma_k$
is given by
\begin{equation}\label{flow}
\partial_t \Gamma_k = 
\frac{1}{2}\, {\rm Tr}\,\frac{1}{\Gamma^{(2)}_k+R}\,\partial_t R\,.
\end{equation}
Here, $t=\ln k$ is the logarithmic scale parameter, Tr denotes a trace
over loop momenta $q$ and a sum over indices and fields, and $R(q^2)$
is the infrared momentum cutoff at momentum scale $k$. $R$ obeys a few
restrictions which ensure that the flow is well-defined and finite
both in the ultraviolet and the infrared. The flow interpolates
between the initial (classical) action in the ultraviolet and the full
quantum effective action in the infrared where the cutoff is
removed. For gauge theories, the flow is amended by a set of modified
Slavnov-Taylor identities.\cite{Litim:1998nf,Freire:2000bq} They
ensure the requirements of gauge symmetry for Green functions in the
physical limit $k\to 0$. Since this approach allows even for
truncations which are non-local in momenta and the fields it is
particularly useful for gauge theories. Furthermore, it is worth
emphasising that the flow only involves fully dressed propagators and
vertices. Therefore, the correct RG scaling properties are represented
by the full flow and truncations with the correct symmetry properties.

So far the flow equation (\ref{flow}) has been applied to Landau gauge
QCD for a determination of the heavy quark effective potential and
effective quark interactions above the confinement
scale.\cite{Ellwanger:1996qf} For an implementation in axial gauges
and further applications in Yang-Mills theories see
Refs.~\cite{Pawlowski:1996ch}.

\section{Infrared regime of QCD}
In quantum field theory, important physical information for low energy
phenomena is given through the momentum structure of Green functions
in the deep infrared regime. In QCD, the characteristic scale
differentiating between strong and weak coupling is $\Lambda_{\rm
QCD}\approx 200$ MeV.  The strongly coupled deep infrared regime is
defined as
\begin{equation}\label{IR1}
p^2 \ll \Lambda_{\rm QCD}^2
\end{equation}
and $p$ denotes a momentum argument of a QCD Green function. As a
consequence of confinement, it is expected that gluon and ghost
propagators display strong deviations from a simple particle pole for
sufficiently small momenta.  Within covariant linear gauges, the
necessary conditions for confinement in terms of local fields were
formulated by Kugo and Ojima.\cite{Kug79} In Landau gauge, they state
that the gluonic correlations are suppressed in the infrared as a
consequence of a mass gap, while the ghost correlations are infrared
enhanced and dominant.  This type of behaviour has already been
detected in solutions of truncated Schwinger-Dyson equations
\cite{Alkofer:2000wg} and stochastic quantisation \cite{Zwa02}, and
lattice simulations \cite{Cucchieri:1997fy}.

Now we turn to the infrared analysis based on (\ref{flow}), where
two-point functions and propagators depend additionally on the cutoff
scale $k$. As long as $k^2\gg p^2$, the propagators barely differ from
the classical ones, given that no quantum fluctuations with momenta
$p$ or smaller have yet been integrated out. In turn, as soon as
\begin{equation}\label{IR2}
k^2\ll p^2
\end{equation}
all quantum fluctuations have been integrated out and physical
quantities as well as general vertex functions upon appropriate
rescaling, are no longer affected by the infrared cutoff.
Consequently, the non-trivial infrared behaviour of physical Green
functions resides in the momentum regime given by (\ref{IR1}) and
(\ref{IR2}).  It can be shown that Green functions in the regime
(\ref{IR1}) depend on the cutoff scale $k$ only parametrically through
dimensionless ratios, or through $k$-dependent renormalisation group
factors that leave the full action invariant.\cite{Pawlowski:2003hq}
This property implies a fixed point behaviour and strongly facilitates
the evaluation of (\ref{flow}) in given truncations.

In order to integrate the flow (\ref{flow}) in the infrared regime, we
introduce an appropriate truncation of $\Gamma_k$ and retain the full
momentum dependence of the QCD propagators. The truncation is amended
by vertices which fulfil the truncated Slavnov-Taylor identities.  The
validity of the truncation has recently been confirmed on the
lattice.\cite{Cucchieri:2004sq} The gluon and ghost two-point function
are parametrised as
\begin{equation}\label{prop}
\Gamma^{(2)}_{k,A/C}(p^2)=z_{A/C} \cdot Z_{A/C}(x)\cdot p^2\,.
\end{equation}
Here, $x=p^2/k^2$, $z$ denotes a possibly $k$-dependent RG factor, and
$Z(x)=x^\kappa(1+\delta Z(x))$ parametrises the non-trivial
modifications of the momentum structure through quantum
fluctuations. The indices $A/C$ refer to the gluon/ghost fields,
respectively. In (\ref{prop}), we have suppressed the trivial colour
structure and the transversal projector for the gluons. The
longitudinal modes do not contribute to the flow in Landau gauge. Note
that (\ref{prop}) is the most general parametrisation of the
propagators and no assumptions have been made upon their structure. 
In
the deep infrared region the behaviour of (\ref{prop}) is dominated by
the term $x^\kappa$, where $\kappa_{A/C}$ is a non-perturbative
anomalous dimension of the gluon and ghost propagator in Landau gauge.
The function $\delta Z$ constitutes the transition behaviour between
the deep infrared regime and the cutoff regime. In addition,
non-renormalisation of the ghost-gluon vertex implies
\cite{Lerche:2002ep}
\begin{equation}
\kappa_A=-2\kappa_C\,,\quad\quad\alpha_s=\frac{g^2}{4\pi}\frac{1}{z_Az_C^2}\,.
\end{equation}
The coefficients $\kappa_{A/C}$ and the value for $\alpha_s$ in the deep
infrared regime are deduced from integrating the flow for the propagators
using (\ref{IR1}) and (\ref{IR2}). This leads to two integral equations for
$\delta Z_{A/C}$ of the form
\begin{equation}\label{int}
\delta Z_{A/C}(x)=F_{A/C}[\delta Z_{A/C},\kappa_{A/C},\alpha_s]\,.
\end{equation}
Explicit expressions for the integrals $F$ are given in
Ref.~\cite{Pawlowski:2003hq}.  The simultaneous solutions of (\ref{int})
lead to explicit solutions for the infrared coefficients $\kappa$,
$\alpha_s$ and $\delta Z$. To leading order, the back-coupling of
$\delta Z$ on the right hand side of (\ref{int}) can be neglected. In
this limit, we find
\begin{equation}\label{result}
\kappa_C=0.59535\cdots\quad \alpha_s=2.9717\cdots
\end{equation}
independently of the cutoff function $R$. Furthermore, the numerical
values agree with the most advanced results obtained within the
Dyson-Schwinger approach and stochastic quantisation.  Iterating
(\ref{int}) beyond leading order ($\delta Z\neq 0$), we find that
$\kappa_C$ varies slightly with the cutoff, ranging between 0.539 for
the sharp cutoff and (\ref{result}) for appropriately optimised
ones. More recently, the infrared region has also been studied in
Ref.~\cite{Kato:2004ry}.

\section{Discussion}
We have shown how, in general, the momentum dependence of Green
functions is extracted in the infrared regime based on an exact
renormalisation group. Applied to QCD in the Landau gauge, we detected
a strong enhancement of the ghost propagator while the gluon
propagator develops a mass gap. Our results provide a further
independent evidence for the Kugo-Ojima scenario of confinement.  In
the simplest possible truncation with dressed propagators our results
match the state-of-the-art within both the Dyson-Schwinger and the
stochastic quantisation approach. This is quite remarkable, in
particular in view of the conceptual and technical differences between
these methods. Current work deals with straightforward extensions of
the present analysis and covers vertex corrections, investigations of
QCD Green functions at finite temperature, and dynamical quarks.  In
either case, the correct RG scaling as well as the inherent finiteness
of the integrated flow are most important.

\end{document}